\documentclass[12pt, onecolumn,draftcls]{IEEEtran}

\usepackage{cite}
\usepackage{latexsym}
\usepackage{epsfig}
\usepackage{verbatim}
\usepackage{amssymb}
\usepackage{graphicx}
\usepackage{color}
\usepackage{epsfig}
\usepackage[cmex10]{amsmath}
\usepackage{stfloats}
\usepackage{setspace}
\usepackage{enumerate}
\usepackage{amsmath,amssymb,bm,amsfonts,bbm}
\usepackage{graphics,graphicx,epsfig,subfigure}
\usepackage{algorithm,algorithmic}
\usepackage{theorem}
\usepackage{threeparttable}
\usepackage{stfloats}
\usepackage{multirow}
\usepackage{mathrsfs}
\usepackage{color}
\usepackage{multirow}
\usepackage{subeqnarray}
\usepackage{cases}
\usepackage{wasysym}
\usepackage{colortbl}
\usepackage[dvipsnames]{xcolor}
\usepackage{url}

\newtheorem{myproof}{Proof}

\newtheorem{mytheorem}{Theorem}

\newtheorem{mydef}{Definition}
\newtheorem{myproperty}{Property}
\newtheorem{mycorollary}{Corollary}

\begin{document}

\title{ Distributed Multi-User Wireless Charging Power Allocation}
\author{Yanju Gu 
}
\maketitle


\begin{abstract}
\noindent
Wireless power charging enables portable devices to be permanently unplugged.
Due to its low transmission power and low transmission efficiency, it requires much longer time slot to charge users compared with that for
data transmission in wireless communication networks.
Besides, each user's demand urgency needs to be taken into consideration for  power allocation.
Therefore, new algorithms are essential for wireless power allocation in multi-user wireless charging networks.
In this paper, this problem is formulated as a
static noncooperative game.
It is shown that there exists a unique Nash equilibrium, which is the static state of the wireless power charging network.
A distributed power allocation algorithm is proposed to compute
the Nash equilibrium of the game.
The main result of the paper consists of rigorous analysis of the  distributed algorithm for power allocation.
The algorithm is shown to
converge to Nash equilibrium of the game with exponentially convergence rate for arbitrary initial value with synchronous scheduling.
Moreover, the distributed algorithm is also convergence guaranteed with asynchronous scheduling under communication delay and packet drops.
Numerical simulations prove the correctness of the analysis and demonstrate the fast convergence of the algorithm and the robustness to synchronous scheduling.

\end{abstract}



\section{Introduction}
Wireless communication aims at providing communication freedom by bringing network to anywhere at anytime.
However, wired charging remains the main way to feed a device battery and
the battery endurance shortage has become the major
bottleneck that hinders the development of ubiquitous wireless
communication systems.
The cables supporting power charging has become the only cables that would require removal.

Wireless power charging
is being developed to
support wireless communication systems.
It is expected that wireless power charging will play a role
in future wireless networks, including the upcoming
Internet of Things systems, wireless sensor networks, RFID networks and small cell networks.
It has the potential of powering devices larger than low-power sensors and tags, e.g., wearable
computing devices and smartphones.
Many leading smartphone manufacturers, such as Apple, Samsung and Huawei, have begun to
release new-generation devices featured with built-in wireless
charging capability.
Driven by the practical needs, wireless power charging has gained momentum from theoretics.
Existing works on wireless charging communication networks can be categorized in two directions.
The first one focuses on exclusive
wireless charging, i.e., wireless power transfer and information
transmission are separated.
The second direction is the research that
wireless charging and information transmission are coupled
to achieve some tradeoff.
In each direction, wireless power charging
has been conducted in various contexts, e.g., point-to-point
channels \cite{ZhangPointToPoint,LimPointToPoint},
relay channels \cite{GuICC,GUTWC},  full duplex channels \cite{DingFullDuplex,ZhongCaiFullDuplex}.

In general, a wireless power charging system consists of a ``transmitter" device connected to a source of power such as mains power lines, which converts the power to a time-varying electromagnetic field,
and one or more ``receiver" devices which receive the power and convert it back to electric power.
From the perspective of transmission range, wireless power charging can be divided into near-field and far-field power charging.
The far-field power charging that utilizes diffused
RF/microwave as a medium to carry radiant energy has a long history and has recently received attention of researchers from communications society.
It is shown that radio frequency signal ambient in the air can be harvested for data transmission, and thus interference signal plays a positive role from the wireless power charging perspective \cite{GuICC}.

On the other hand,
for the near-field wireless power charging,
traditionally,
transmission range of centimeters has been realized based on the principle of electromagnetic inductive coupling.
For instance, in the
Qi standard, the distance between a receiver, e.g., a mobile
phone, and its power transmitter should not exceed $4$ cm \cite{Qi}.
More recently,
to boost the efficiency of power charging,
magnetic resonant coupling (MRC) has been extensively studied which has higher power charging efficiency as well as longer operation range up to a couple of meters.
MRC effectively avoids power leakage to non-resonant externalities and thus ensures safety to the neighboring environment.
In \cite{MMC-Dina}, an experiment system with multiple transmitters is successfully delivered which can simultaneously charge a cell phone at distances
of about half meter, and work independently how the phone is oriented in user's pocket and
in \cite{Katabi-Multi} the authors further prototyped charging, simultaneously, multiple devices including smart phones
and tablets.


No matter for which kinds of wireless power charging technology, the number of power transmitter is expected to be small and each transmitter needs to serve multiple users at the same time.
Thus how to allocate the charging power to different users needs to be studied and is still missing.
Different from the multi-user communication, multi-user wireless power charging has its unique feature and raise new challenges.
First, wireless power charging, due to low transmission power and low transmission efficiency, takes much longer time to full charge users compared with
data transmission in wireless communication networks.
Besides,
 the battery capacities for each user
may vary and lead to different levels of aspiration for power
charging.
Therefore, with limited  power at the transmitter,
multi-user charging power allocation is an important issue to be solved with consideration of different users expected amount and duration.

In this paper, we model the competition of limited charging power among users via game theory.
In general, the priority of power allocation should give to a user with low state of charge and tight charging deadline.
In order to be able to ensure at which state the multi-user wireless charing networks will effectively operate, we provide rigourous analysis of the Nash  equilibrium.
The contributions can be summarized as follows
\begin{itemize}
\item
We propose a game framework for wireless power allocation to multiple users with different charging needs and deadlines.
Users compete for the limited power by independently submitting a unit price bid to the power transmitter, and proportional sharing
of power follows for the power allocation.
\item
We analyze this wireless power allocation problem using rigorous game-theoretic analysis.
We show the existence and uniqueness of the Nash equilibrium for wireless power allocation.
Besides, we show for arbitrary initial values the convergence to the Nash equilibrium with exponential rate.
\item
In many practical scenarios, the inter-user
information exchange is asynchronous since random data packet losses may occur,
and different nodes may update at different frequencies.
These impacts are considered in computing the Nash equilibrium, which represent a general framework including the synchronous updating as a special case.
\item
The  power allocation algorithm takes the advantage of distributed computation \cite{du2014distributed, du2013network, asilomar}.
The users' charging capacity and charging deadline is not required to be shared while computing the Nash equilibrium.
\end{itemize}


The rest of this paper is organized as follows.
Section II gives the problem modeling and formulation.
Section III rigourously shows the existence and uniqueness of the Nash equilibrium,  proposes a distributed algorithm to reach the Nash equilibrium with exponentially convergence rate, which is also robust to communication delay and packet loss.
Section IV shows the simulation results of distributed multi-user wireless charging power allocation.
Finally, Section V concludes the
paper.

%

\section{System Model and Problem Formulation}\label{Model}
We consider a wireless charging server with total transition power $P$, which serves the charging services of $M$  users in the its effective charging area.
$\mathcal{V}=\{1,\ldots, M\}$ denotes the set of users.
Similar
to the single-input-multiple-output  communication systems, the transmitter's power is wirelessly transferred to multiple users.
Let $h_i$ be the wireless power charging  efficiency to user $i$, with 
value varying from 0 to 1,
which is defined as the received power at the receiver
divided by the  input power for charing $i$ at the charging server.
Note that with different types of wireless power transmission schemes, $h_i$ has different physical meaning.
More specifically, in the MRC based wireless power charging system,
$h_i$
depends on the mutual inductance between electromagnetic coils
of the transmitter and $i^{th}$ receiver \cite{MMC-Dina}.
While in the RF  based wireless power transmission schemes, $h_i$ is the product of the transmit efficiency,
the free-space propagation efficiency and the receive
efficiency \cite{GUTWC}.
In the state-of-the-art, in order to improve the end-to-end
power charging efficiency, researchers¡¯ efforts were focused on
enhancing the wireless power transmission efficiency via designing the non-radiated
magnetic field or RF signal array
to reduce the side lobes of the beam
pattern while keep
its main lobe keeps spillover losses to a minimum \cite{Katabi-Multi}.
In the RF based wireless power transmission schemed, attaining higher receive efficiency was also attempted
through the design of high-performance rectifying antennas \cite{efficiency}
(i.e. rectennas), which convert the incident RF power back to
DC.

In contrast to the high data transmission rate achieved in wireless communication networks,
to transmit certain amount of power for wireless charging takes much longer time due to the limited of wireless transmission power and low wireless charging effiency.
Therefore, each user $i$ has an expected charging period $[0, D_i]$, after $D_i$ user $i$ has to leave the effective charging area, and $C_i$ is the amount of energy that is  needed to fulfill $i$'s power consumption, such as communication, requirement   at current time.
Each time, user $i$ submits a unit price bid $x_i$ to compete for the amount of power transmitted from the charging server,
and proportional sharing  mechanism \cite{ProptoSharing} applies, i.e., $P$ is allocated to each user in proportion to their unit price bids.
Hence, the received power at user $i$ is
\begin{equation}\label{power}
y_i = \frac{x_iP}{\sum_{j\in \mathcal{V}} x_j}h_i.
\end{equation}
Note that a reasonable  unit price bid should be non-negative
i.e.,
\begin{equation}\label{set}
\mathcal S_i=\{ x_i>0\}.
\end{equation}
The user's satisfaction to the wireless charging service relies on whether  the charged device can fulfil its requirement.
Hence, the satisfaction metric of each user can be defined as
\begin{equation}\label{urgency}
 s_i =   \frac{y_i}{C_i/D_i}.
\end{equation}
Note that $C_i/D_i$ means the received amount of power that is able to full charge user $i$, which
is the expected minimum  charging rate for user $i$.
Thus, $s_i$
is the ratio of the allocated power charging rate $ y_i$ with the minimum expected charging rate.
The physical meaning is that if $s_i= 1$,  user $i$ is charged with $C_i$ within exactly charging period $D_i$;
if $s_i>1$, it denotes that user $i$ will full fill the charge of $C_i$ power within the charging time $D_i$;
and if $s_i<1$, the charging power rate $x_i$ can not satisfy user $i$'s power charging requirement within $[0, D_i]$ charging period.

In order to obtain energy from the charging server,  user $i$ needs to pay for the service.
Therefore, the utility function that user $i$
 tries to maximize can be formulated as
\begin{equation}\label{utility0}
\mathcal{U}_i(x_i, \bm x_{\mathcal{V}\setminus i})
=
s_i -
\lambda x_i \frac{P x_i}{ \sum_{j\in \mathcal{V}} x_j}, \quad i\in \mathcal V,
\end{equation}
where
$x_i\frac{P x_i}{ \sum_{j\in \mathcal{V}} x_j}$
denotes the payment required to obtain $\frac{P x_i}{ \sum_{j\in \mathcal{V}} x_j}$ amount of power,
and $\lambda>0 $ implies how to balance between the satisfaction of charging power rate and the payment.
The utility function $\mathcal{U}_i(x_i, \bm x_{\mathcal{V}\setminus i})$
is a measure of user $i$'s preference expressed 
by the amount of satisfaction he or she receives minus the amount of he/she pays
Note that if the urgency  is not take into consideration, i.e., $\lambda = +\infty$, users would bid as low as possible to reduce their payments.
Substituting (\ref{power}) and (\ref{urgency}) into (\ref{utility0}), the utility function is
\begin{equation}\label{utility}
\mathcal{U}_i(x_i, \bm x_{\mathcal{V}\setminus i}) =\frac{D_i P h_ix_i}{C_i\sum_{j\in \mathcal{V}} x_j}  -
\lambda \frac{P x^{2}_i}{ \sum_{j\in \mathcal{V}} x_j}, \quad i\in \mathcal V.
\end{equation}
The multi-user wireless charging problems can then be formalized as a non-cooperative game with the following triplet structure:
\begin{equation}\label{game}
\mathcal{G}=\langle\mathcal{V},
\{\mathcal{S}_i\}_{i\in \mathcal{V}},
\{\mathcal{U}_i(x_i, \bm x_{\mathcal{V}\setminus i})\}_{i\in \mathcal V}\rangle,
\end{equation}
where $\mathcal{V} $ denotes the game players;
$\mathcal{S}_i  $ being the range of $x_i$ represents the game strategy, which is a subset of positive real numbers;
and
$\{\mathcal{U}_i(x_i, \bm x_{\mathcal{V}\setminus i})\}_{i\in \mathcal V}$ is the utility function for player $i$.
The concept of a best response is central for problem formulation and analysis with game theory.
The formal definition is given as follows.

\begin{mydef}\label{def}
The best response function
$\mathcal F_i(\bm x_{\mathcal{V}\setminus i})$
of user $i$ is the best strategy for user $i$ given the power allocation
of the others, which
is then the solution to the following maximization
problem:
\begin{equation}\label{response}
\mathcal F_i(\bm x_{\mathcal{V}\setminus i})
=
\text{arg}\,\underset{x_i}{\text{max}}
\
\mathcal{U}_i(x_i, \bm x_{\mathcal{V}\setminus i}).
\end{equation}
Note that for each $i$, the maximum is taken over $x_i$ for a fixed
$\bm x_{\mathcal{V}\setminus i} \triangleq [x_1,\ldots, x_{i-1}, x_{i+1}, \ldots, x_{M}]^T$.
\end{mydef}

We show in Appendix \ref{A} that
the best response function for user $i$ in analytically form is
\begin{equation}\label{BR0}
\mathcal F_i(\bm x_{\mathcal{V}\setminus i})
=
\big[(\sum_{j\in \mathcal{V}\setminus i}x_j)^2  +  K_i\sum_{j\in \mathcal{V}\setminus i}x_j \big]^{\frac{1}{2}} - \sum_{j\in \mathcal{V}\setminus i}x_j,
\end{equation}
where $K_i = \frac{D_i h_i}{\lambda C_i}$.
The best response function (\ref{BR0}) reflects the fact that all users take independently their best available actions
  in order to pursue
their own individual objectives as expressed by the particular
choice of $x_i$.
Since $\sum_{j\in \mathcal{V}\setminus i}x_j> 0$, the best response function (\ref{BR0}) can be reformulated as
\begin{equation}\label{BR2}
\mathcal F_i(\bm x_{\mathcal{V}\setminus i})
=
{K_i}\big[
\big(1+\frac{K_i}{\sum_{j\in \mathcal{V}\setminus i}x_j}
\big)^{1/2}
+1\big]^{-1},
\end{equation}
and it is evident that when some user $j\neq i$ increases, $\mathcal F_i(\bm x_{\mathcal{V}\setminus i})$ will decrease.
This observation indicates
that what is best for one link depends in general upon actions
of other links due to the total energy constraint at the charging server.
Thus, an equilibrium
for the whole system is reached when every user is unilaterally
optimum, i.e., when, given the current strategies of the others,
any change in his own strategy would result in a in terms of $\mathcal F_i(\bm x_{\mathcal{V}\setminus i})$.
This
equilibrium constitutes the celebrated notion of Nash equilibrium.
We define
the joint best response function as $\mathcal{F} \triangleq \{\mathcal{F}_1(\bm x_{\mathcal{V}\setminus 1}),\ldots, \mathcal{F}_M(\bm x_{\mathcal{V}\setminus M}) \}$
and
$\mathcal S = \mathcal{S}_1\times \ldots, \times  \mathcal{S}_M$,
which are a Cartesian product of $\mathcal{F}_i(\bm x_{\mathcal{V}\setminus i})$
and $\mathcal{S}_1,\ldots, \mathcal{S}_M$, respectively.
Then the corresponding Nash equilibrium for the multi-user wireless charging power allocation  game
is formally defined as follows.
\begin{mydef}
A pure strategy profile $\bm x^{\ast} =[x_1^{\ast}, \ldots, x_M^{\ast}]\in \mathcal{S}$ is a Nash equilibrium of the game $\mathcal{G}$ in (\ref{game}) if $\bm x^{\ast}= \mathcal{F}(\bm x^{\ast})$.
\end{mydef}

In general, the game $\mathcal G$ in (\ref{game}) may admit multiple equilibria.
In the next section, we will prove the uniqueness of the Nash equilibrium and we analyze the property  of how to converge to such an equilibrium in a fully distributed way \cite{}.

\section{Nash Equilibrium and Distributed Algorithm Analysis }\label{Nash}
\subsection{Nash Equilibrium Analysis}
In this subsection, we will show that there is always a unique Nash equilibrium for the game $\mathcal G$ in (\ref{game}) is always exists.
Note that the standard interference functions introduced by Yates \cite{Yates} have been very influential on the Nash equilibrium analysis
and design of distributed power control laws for wireless communication networks.
Quite a number of works \cite{Richard-Standard, Chen-Standard,Sun-Standard} has been motivated by \cite{Yates} which conduct the Nash equilibrium analysis via
showing that the corresponding best response function is a  standard function.
While the standard function framework is powerful, the framework does not shown the existence of fixed-points and there is no   guarantees on the rate of convergence of the iterates.
Note that, recently, \cite{duJMLR, informationmatrix,pairwise} shows that the distributed belief propagation algorithm equals maximizing local utility function to find a Nash equilibrium and the convergence rate is analyzed.
Further, the convergence rate of a variant of standard belief propagation is also analyzed in \cite{du2017proactive}.
However, these conclusions cannot be used in our problem due to the fact that the type of utility functions are different.
Motivated by \cite{Yates} and \cite{duJMLR, informationmatrix,pairwise, du2017proactive}, we  show some unique properties of the best response function in (\ref{BR0}) first and then conduct the Nash equilibria and distributed computation algorithm analysis.

\begin{myproperty} \label{P_FUN}
 The updating operator $\mathcal{F}(\cdot)$ satisfies the following properties:

\noindent
P 1.1: 
For arbitrary $\bm x, \bm y \in \mathcal S$,
 $\bm x\leq \bm y$ implies
 $\mathcal{F}(\bm{x}) \leq  \mathcal{F}(\bm{y})$.

\noindent
P 1.2  For arbitrary $\bm x \in \mathcal S$ and $\alpha<1$, $\mathcal{F}(\cdot)$ satisfies
$\mathcal{F}(\alpha\bm{x}) >  \alpha\mathcal{F}(\bm{x})$
and
$\alpha^{-1}\mathcal{F}(\bm{x}) >
\mathcal{F}(\alpha^{-1} \bm{x})$.
\end{myproperty}
\begin{myproof}
The proof is shown in Appendix \ref{B}.
\end{myproof}
Note that Nash equilibrium does not always exist.
If the set $\mathcal S_i$ for all $i\in \mathcal V$ was a closed, bounded, and
convex subset of the one dimensional Euclidean space, 
and 
$\mathcal{U}_i(\bm x_{\mathcal{V}\setminus i})$  is a strictly concave function in $ \mathcal S$, then
 the noncooperative game (\ref{game})
has a Nash equilibrium \cite[p. 173]{DynamicGame}.
This theorem is often used to check the existenceness of Nash equilibirum in literature \cite{Kucera,gametutorial,exitence}.
However, $\mathcal S_i$ is not a compact set as shown in (\ref{set}), and traditional analysis method fails.
Next the existence  of Nash equilibria is studied.

\begin{mytheorem} \label{exist}
The  non-cooperative wireless charging game in (\ref{game}) admits at least one Nash equilibrium, i.e.,
$\bm x^{\ast}= \mathcal{F}(\bm x^{\ast})$.
\end{mytheorem}
\begin{myproof}
Since $\mathcal F(\cdot)$ is a continous function and considering P 1.1, $ \mathcal F(\bm x^{n})$ must be upper bounded by
$\mathcal F(+\infty\bm 1)$ if
$
\lim_{\bm x_{\mathcal{V}\setminus i}\to \infty }F_i(+\infty\bm 1)$ exists.
By substituting $
\bm x^{(n-1)}_{\mathcal{V}\setminus i}=+\infty\bm 1$
into $\mathcal{F}(\cdot )$ in (\ref{BR2}), we have
\begin{eqnarray}\label{lim}
u_i
&=&\lim_{\bm x_{\mathcal{V}\setminus i}\to \infty }
\frac{ K_i }
{ \big[(1 + K_i \frac{1}{\sum_{j\in \mathcal{V}\setminus i}x_j^{(n-1)}} \big]^{\frac{1}{2}} + 1 }\nonumber\\
&=&\frac{1}{2}K_i.
\end{eqnarray}
By defining $\bm u = \frac{1}{2}\bm K
=\frac{1}{2}[K_1,\dots K_i, \dots K_M]^T
$.
we have $\mathcal F( \bm x)<\bm u$ for all $\bm x>\bm 0$.
Then we choose arbitrary $\bm x^{(0)}>\frac{1}{2}\bm K$, we have 
$\bm x^{(1)}= \mathcal F(\bm x^{(0)})<\bm u$.
By repeatly applying $\mathcal F(\cdot)$ on both sides of the inequality we have $\bm x^{(n)}<\bm x^{(n-1)}$.
Therefore  $\bm x^{\{n\}}$ is a monotonically decreasing sequence lower bounded by $\bm 0$. 
According to monotonic theorem, it must converge.
Thus there exsit a fixed point $\bm x^{\ast}$ which satifies 
$\bm x^{\ast} = \mathcal F(\bm x^{\ast})$.
Thus the non-cooperative power game in (\ref{game}) admits at least one Nash equilibrium, i.e.,
$x_i^{\ast} = \mathcal{F}_i(\bm x^{\ast}_{\mathcal{V}\setminus i}) $ for all
$i\in \mathcal V$.
\end{myproof}

It is possible that a game has multiple Nash equilibirums.
Next, we  show the uniqueness of the Nash equilibrium in (\ref{game}).
\begin{mytheorem} \label{unique}
The  Nash equilibrium for the non-cooperative wireless charging game in (\ref{game}) is unique.
\end{mytheorem}
\begin{myproof}
The uniqueness property of the converged $\bm x^{(n)}$, denoted as $\bm x^{\ast}$ is proved by contradiction.
Suppose there are two  distinct fixed points $\bm x^{\ast}$ and $\tilde{\bm x}^{\ast} $.
Without loss of generality, there exist $i\in \mathcal V$ such that
$\tilde{x}_i^{\ast}< x_i^{\ast}$.
It is clear that, there must exist $0<\alpha_i<1 $ such that
$  \tilde{x}_i^{\ast}=\alpha_i x_i^{\ast}$.
Let $\alpha$ denote the
the smallest $\alpha_i$ for all $i\in \mathcal V$, we have
\begin{equation}\label{inequ}
\alpha \bm x^{\ast}  \leq \tilde{\bm x}^{\ast}, \quad \alpha<1.
\end{equation}
Applying $\mathcal F(\cdot)$ on (\ref{inequ}), and considering P 1.1, we have
\begin{equation} \label{inequ1}
 \mathcal F( \alpha\bm x^{\ast})
\leq \mathcal F( \tilde{\bm x}^{\ast}) =\tilde{\bm x}^{\ast},
\end{equation}
where the  equality follows the definition of fixed point of $\bm x^{\ast}$.
Besides, as
$x_j^{\ast}> 0$ and $0<\alpha<1$,
following P 1.2, we have
$\mathcal F( \alpha\bm x^{\ast})> \alpha\bm x^{\ast}$.
According to (\ref{inequ1}), we obtain
\begin{equation}\label{contra2.1}
\tilde{\bm x}^{\ast}> \alpha\bm x^{\ast}.
\end{equation}
Now we can conclude that (\ref{inequ}) and (\ref{contra2.1}) is a contradiction, and therefore
$\bm x^{\ast}$ and $\tilde{\bm x}^{\ast} $ are the same fixed point.
Therefore, $\bm {x}^{(n)}$ converges to a unique fixed positive value.
\end{myproof}

\subsection{Distributed Algorithm and Convergence Analysis}
In the previous subsection,
the existence and uniqueness of the Nash equilibrium have been established by Theorem \ref{exist} and Theorem \ref{unique}, respectively.
In order to compute the unique Nash equilibrium, the charging server could gather all the users' information together to perform the computation.
However, the parameter $K_i$, which depends on user $i$'s charging deadline and capacity,
is the privacy information of user $i$, and individual user may not be willing to transmit  these information to the charging server.
In order to reach the Nash equilibrium of the game without sharing individual privacy information,
distributed iterative computation procedure is preferred.
In this section, we first give the synchronous updating algorithm and then analytically prove that  for arbitrary initial value of the distributed computation algorithm, the distance between initial value and the Nash equilibrium decreases exponentially.
We will also study how to choose the initial value
to make the distributed algorithm  converges faster.

We introduce some preliminary definitions
to provide a formal description of the distributed computation algorithm.
We assume,
without any loss of generality, that the set of times at which one
or more users update their strategies is the discrete set
$\mathcal{T}=\{0, 1, 2, \ldots\}$.
Let $x_i^{(n)}$ with $n\in \mathcal{T}$ denote the unit price bid of user $i$ at the $n^{\textrm{th}}$ iteration.
Hence, according to
(\ref{BR0}), the unit price bid of user $i$
is given by

\begin{equation}\label{evolution}
\begin{split}
x_i^{(n)}
& =\mathcal{F}_i(\bm x^{(n-1)}_{\mathcal{V}\setminus i})\\
&=\big[(\sum_{j\in \mathcal{V}\setminus i}x_j^{(n-1)})^2  +
 K_i\sum_{j\in \mathcal{V}\setminus i}x_j^{(n-1)} \big]^{\frac{1}{2}}
- \sum_{j\in \mathcal{V}\setminus i}x_j^{(n-1)}
\end{split}
\end{equation}
It is noteworthy that user $i$ only needs to compute $x_i^{(n)}$ with the updated $x_j^{(n-1)}$ where $j\neq i$, without user $j$ sharing the privacy parameter $K_j$.
We summarize the distributed iterative algorithm as follows.
The algorithm is started by setting user $i$'s bid as $x^{(0)}_{i}$.
Each node $i$  computes its updated message according to (\ref{evolution})  at independent time $n\in\mathcal{T}$ with its available
$\bm x^{(n-1)}_{\mathcal{V}\setminus i}$.
Then the joint distributed computation function is
\begin{equation}\label{fixed}
\bm x^{(n)} = \mathcal{F}(\bm x^{(n-1)} )
 \triangleq \{\mathcal{F}_1(\bm x^{(n-1)}_{\mathcal{V}\setminus 1}),\ldots, \mathcal{F}_M(\bm x^{(n-1)}_{\mathcal{V}\setminus M}) \}.
\end{equation}
Mathematically speaking, analyzing the Nash equilibrium of (\ref{game}) equals to analyze the corresponding fixed point problem of (\ref{fixed}) \cite{Nash}.
It is shown in \cite{duJMLR, informationmatrix} that if an iterative function has some specific properties, the convergence is guaranteed and the convergence rate can be identified.
We next show the convergence property of $\mathcal{F}(\cdot)$ follow the proof procedure in \cite{duJMLR, informationmatrix}.

\begin{mytheorem} \label{C-converge-iff}
$\bm x^{(n)}$  converges to the unique Nash equilibrium $\bm x^{\ast}$ of the non-cooperative wireless charging game in (\ref{game})  for any initial bids $ \bm x^{(0)}\in \mathcal S$.
\end{mytheorem}
\begin{myproof}
As shown in Theorem \ref{unique}, the Nash equilibrium is unique.
Then the posiibility of inital bid $\bm x^{(0)}$
 can be categorized into three groups:
1) the bid for each user is larger than its Nash euilibrium, i.e.,  $\textbf {x}^{\ast}<\textbf x^{(0)}$.
2) the bid for each user is smaller than its Nash euilibrium, i.e.,  $\textbf x^{(0)}<\textbf {x}^{\ast}$.
3) some of  the bid for each user is larger than its Nash euilibrium, and some of the bid for each user is smaller than its Nash euilibrium.

\noindent 1)  $\textbf {x}^{\ast}<\textbf x^{(0)}$:
There exist $\alpha\geq 1$ which satisfies
$\textbf x^{(0)}\leq \alpha \textbf {x}^{\ast} $.
Following the monotinic property, we have $\textbf{x}^{\ast}=\mathcal F(\textbf {x}^{\ast})<\mathcal F(\textbf x^{(0)}) <\mathcal F(\alpha\textbf x^{\ast})
\leq \alpha\mathcal F(\textbf x^{\ast})=
\alpha \textbf x^{\ast} $.
Let sequence $\tilde{\textbf x}^{(0)} = \alpha \textbf x^{\ast}$ and
$\tilde{\textbf x}^{(1)} = \mathcal F(\tilde{\textbf x}^{(0)})
$
Then we define a new sequence $\{\tilde{\textbf x}^{(t)}\}$.
As $\tilde{\textbf x}^{(1)}=\mathcal F(\tilde{\textbf x}^{(0)})
=\mathcal F(\alpha \textbf x^{\ast})\leq \alpha \mathcal F( \textbf x^{\ast}) =\tilde{\textbf x}^{(0)}$.
Then, $\tilde{\textbf x}^{(1)}< \tilde{\textbf x}^{(0)}$ implies that $\tilde{\textbf x}^{(t)}$ is a monotinocally decreasing sequence.
Thus $\tilde{\textbf x}^{(t)}$ converges.
Since $\textbf x^{\ast}$ is the unique fixed point, $\tilde{\textbf x}^{(t)}$ also converges to $\textbf x^{\ast}$.
As $\textbf x^{\ast} < \textbf x^{(0)} <\tilde{\textbf x}^{(0)}$,
$\mathcal F(\textbf x^{\ast}) < \mathcal F(\textbf x^{(0)}) <\mathcal F(\tilde{\textbf x}^{(0)})$.
Thus
$\textbf x^{(0)} $ with $\textbf x^{(0)}>\textbf x^{\ast}$, must converges to $\textbf x^{\ast} $.

\noindent 2) $\textbf 0<\textbf x^{(0)}<\textbf {x}^{\ast}$:
There exist $1>\beta>0$ which satisfies $ \tilde{\textbf x}^{(0)}
=  \beta \textbf x^{(\ast)} < \textbf x^{(0)}$.
$\tilde{\textbf x}^{(1)}
=\mathcal F(\tilde{\textbf x}^{(0)})= \mathcal F(\beta\textbf x^{\ast})
\geq \beta\mathcal F(\textbf x^{\ast})
=
\beta \textbf x^{\ast}
=\tilde{\textbf x}^{(0)}$.
Thus $ \textbf x^{(0)}$ converges to $ \textbf x^{(\ast)}$ with
$\textbf 0<\textbf x^{(0)}<\textbf {x}^{\ast}$.

\noindent 3) indefinite:
There exist $\textbf x_{u}$ and  $\textbf x_{l}$ such that
$\textbf 0<\textbf x_{l}<\textbf x^{(0)}<\textbf x_{u}$
and
$\textbf 0<\textbf x_{l}<\textbf x^{\ast}<\textbf x_{u}$.
Follwing case 1) $\lim_{\ell\to \infty } \mathcal F^{(\ell)}(\textbf x_{u}) =
\textbf x^{\ast}$;
and
Follwing case 2) $\lim_{\ell\to \infty } \mathcal F^{(\ell)}(\textbf x_{u}) =
\textbf x^{\ast}$;
Thus, $\lim_{\ell\to \infty } \mathcal F^{(\ell)}(\textbf x_{u})
\leq \lim_{\ell\to \infty } \mathcal F^{(\ell)}(\textbf x^{(0)})
\leq \lim_{\ell\to \infty } \mathcal F^{(\ell)}(\textbf x_{u}) $.

\end{myproof}

 Another fundamental question is how fast the convergence would be.
Define a set
\begin{equation}
\begin{split}
\mathcal C
 =&\{\bm x^{(n)}|  \bm x_{\mathcal{U}} \geq \bm x^{(n)} \geq \bm x^{\ast}+ \epsilon \textbf{1}\} \\
&\cup
 \{\bm x^{(n)}|
\bm x^{\ast}- \epsilon \textbf{I} \geq \bm x^{(n)} \geq \bm x_{\mathcal{L}}\},
\end{split}
\end{equation}
with $\epsilon>0$ being an arbitrary small value.
Next, we give the definition of the part metric first, and then
show that the distance between  $\{\bm x^{(n)}\}_{n=1,\ldots}$ and the unique Nash equilibrium $\bm x^{\ast}$ decreases  doubly exponentially fast with respect to part metric in $\mathcal C$.

For a proof
of the fact that $\mathrm{d} (x, y)$ is a metric and for other properties of the part metric
we refer to \cite{PartMonotone}.
Then following the proof in \cite{duJMLR}, we can show the convergence rate is doubly exponential.
\begin{mytheorem}\label{RateCov}
With the initial bid $\bm x^{(0)}$ set to be arbitrary  value in $\mathcal S$,
 the distance between $\bm x^{(n)}$ and  $\bm x^{\ast}$ decreases doubly exponentially fast with $n$ increases.
\end{mytheorem}

The physical intuition of Theorem \ref{RateCov} is that the sequence $\{\bm x^{(n)}\}_{l=1,...}$ converges at a geometric rate before $\bm x^{(n)}$ enters $\bm x^{\ast}$'s neighborhood, which can be chose arbitrarily small.
Next we show how to
make sure the monotonic convergence.

\begin{mytheorem} \label{C-converge-iff}
$\bm x^{(n)}$ converges monotonically  for any initial value $ \bm x^{(0)}>\bm 0$  if
the function $\mathcal{F}(\cdot)$ satisfies
$\{\bm x\geq \mathcal{F}(\bm x)
\cup
\mathcal{F}(\bm x)\geq \bm x \} \neq \emptyset$.
\end{mytheorem}
\begin{myproof}
On one hand, if $\bm x \geq \mathcal F(\bm x)$ holds for some $\bm x $, we can establish that
${\bm{ x }}^{(0)}\geq \mathcal F({\bm{ x }}^{(0)})$ or equivalently,
${\bm{ x }}^{(0)}\geq {\bm{ x }}^{(1)}$.
By induction, this relationship can be extended to
${\bm{ x }}^{(0)}\geq {\bm{ x }}^{(1)}\ldots
\geq {\bm{ x }}^{(n)}$.
Besides, if $\mathcal F(\bm x)  \geq \bm x$ holds, it implies
$\ldots \geq {\bm{ x }}^{(n)}\geq  \ldots \geq {\bm{ x }}^{(1)} \geq
{\bm{ x }}^{(0)}$.
Therefore, $\{\bm x|\bm x \geq  \mathcal F(\bm x)\cup  \mathcal F(\bm x)  \geq \bm x \} \neq \emptyset$ implies
${\bm{ x }}^{(0)}\ldots   {\bm x}^{(n)}
$ must be a monotonic sequence.
Since ${\bm x}^{(n)}> \bm 0$, for all $ n\geq 0$ and considering P 1.1,
we have
$\mathcal F(\bm x)$ must within the range of
$\mathcal F(\bm 0)$ and
$\mathcal F(+\infty\bm 1)$.
As shown in (\ref{lim}), we have $\mathcal F(+\infty\bm 1)= \bm u$.
On the other hand,
substituting $\bm x^{(n-1)}_{\mathcal{V}\setminus i}=\bm 0$ into (\ref{evolution}), we obtain
$\mathcal F(\bm 0)=\bm 0$.
Hence, the bounded monotonic
$\bm x^{(n)}$ converges.
\end{myproof}

Note that to guarantee
 $\{\bm x\geq \mathcal{F}(\bm x)
\cup
\mathcal{F}(\bm x)\geq \bm x \} \neq \emptyset$ we need to choose ${\bm{ x }}^{(0)}$ such that
 ${ \bm x}^{(0)}\geq {\bm{ x }}^{(1)}
$ or
${\bm{ x }}^{(1)}\geq {\bm{ x }}^{(0)}
$.
Also, if  ${\bm{ x }}^{(0)}\geq {\bm{ x }}^{(1)}
$ or
${\bm{ x }}^{(1)}\geq {\bm{ x }}^{(0)}
$,
we have  $\{\bm x\geq \mathcal{F}(\bm x)
\cup
\mathcal{F}(\bm x)\geq \bm x \} \neq \emptyset$.
Thereby, Theorem \ref{C-converge-iff}
is equivalent to
$\bm x^{(n)}$ converges if
${\bm{ x }}^{(0)}\geq {\bm{ x }}^{(1)}
$ or
${\bm{ x }}^{(1)}\geq {\bm{ x }}^{(0)}
$
with
${\bm{ x }}^{(0)}>0$.
Hence, to guarantee the convergence in a monotonic fashion, we can simply set ${\bm{ x }}^{(0)}>[\bm u, \bm x_{\mathcal U}) $ if $\bm u\geq \bm x_{\mathcal U}$.
Next, we will show how to choose the initial value
${\bm{ x }}^{(0)}$ to make ${\bm{ x }}^{(n)}$ converges faster in a monotonic fashion.

\begin{mytheorem}\label{Coro1}
 $\bm x^{(n)}$ is convergence guaranteed with  arbitrary initial value
$\bm x^{(0)} $ within the set
$[\bm u, \bm x_{\mathcal U})$,
and with different $\bm x^{(0)} $ in this range, $\bm x^{(n)} $ converges to the same value.
With $\bm x^{(0)}=  \bm u$,  the iteration (\ref{evolution}) converges faster than the same iteration with any other
$\bm x^{(0)}
\in (\bm u, +\infty)$.
\end{mytheorem}
\begin{myproof}
We denote by $\bm x^{(n)}$ the vector sequence of (\ref{evolution}) with initial value $\bm x^{(0)}=  \bm u$ and by $\bm y^{(n)}$ the vector sequence with $\bm y^{(0)}
\in (\bm u,+\infty)$.
We shall prove that
\begin{equation}\label{target}
\bm x^{(n)} < \bm y^{(n)} \quad \textrm{and} \quad ||\bm x^{(n)}-\bm x^{\ast}|| < ||\bm y^{(n)}-\bm y^{\ast}||
\end{equation}
for $n = 0,1,\ldots,$.
We prove the above results by induction.
At the start, we have
$\bm x^{(0)}=  \bm u<\bm y^{(0)}$.
Assume that $\bm x^{(n)}<\bm y^{(n)}$.
According to P 1.1,  we obtain
$\bm x^{\ast}\leq \bm x^{(n+1)}= \mathcal F(\bm x^{(n)})<\bm y^{(n+1)}=\mathcal F(\bm y^{(n+1)})$.
Thus, (\ref{target}) is proved and
with $\bm x^{(0)}=  \bm u$, then the iteration (\ref{evolution}) converges faster than the same iteration with any other
$\bm x^{(0)}
\in [\bm u,+\infty)$.
\end{myproof}

We have analyzed the convergence property of synchronous updating $x_i^{(n)}$ for all $i\in \mathcal V$.
The synchronous updating can be used in centralized computation, where all the information including $K_i$ for all 
$i\in \mathcal V$ are gathered in one computation centre.
However, from privacy preserving perspective, users  are averse from
sharing their
charging capacity and charging deadline.
The distributed power allocation algorithm takes the advantage of privacy preserving.
In distribueed algorithm, 
the updating of $x_i^{(n)}$ at each iteration only occurs after  every user
receiving updated $x_i^{(n-1)}$ from all its neighbors.
However, in practice networks,  due to communication packet drops or processing delay the convergence speed of the distributed algorithm would be slow.
Next, a totally asynchronous bid updating is analyzed.

\subsection{Asynchronous Power Allocation}\label{simulation}
With the totally asynchronous updating scheme, all the users compute
their individual best response  function (\ref{BR0}) in a totally asynchronous way. More specifically,
some users are allowed to update (\ref{BR0}) more
frequently than others with outdated information about the updated bid from the others.
For example,
when node $i$ computes $x_{i}^{(n)}$,
it may only have $\bm x_{j}^{(s)}$ computed by other node $j$ with
$s\leq n-1$.
In order to capture these asynchronous properties of message exchanges, totally asynchronous updating has been adopted in wireless communication networks \cite{AsynNetwork,clockconf,}.
Next we introduce the totally asynchronous scheduling definition and formulate the totally asynchronous distributed  best response updating.

\begin{mydef} \label{def asyn}
Totally Asynchronous Scheduling:
Let the bid
 available
to node $i$ from user $j$ at time $n$ are
$x_j^{\tau_i(n-1)}$, with $\tau_i(n-1)$ satisfying  $0 \le \tau_i(n-1) \le n$ and $\mathop {\lim }\limits_{n \to \infty } \;\tau_i(n) = \infty$ for all $i=1,..., M$.
\end{mydef}

According to Definition \ref{def asyn}, the asynchronous updating version of (\ref{evolution})
can be expressed as
\begin{eqnarray}\label{evolution-asyn}
x_i^{(n)}
&=&\mathcal{F}_i(\bm x^{\tau_i(n-1)}_{\mathcal{V}\setminus i})\nonumber\\
&=&
\big[(\sum_{j\in \mathcal{V}\setminus i}x_j^{\tau_i(n-1)})^2  +  K_i\sum_{j\in \mathcal{V}\setminus i}x_j^{\tau_i(n-1)} \big]^{\frac{1}{2}}\nonumber\\
&& - \sum_{j\in \mathcal{V}\setminus i}x_j^{\tau_i(n-1)}.
\end{eqnarray}
Note that $x_i^{(n)}$ is the outgoing information from user $i$, the corresponding avaiable bid used for computation at user $j$ is
denoted by
$x_i^{\tau_j(n-1)}$.
and $\bm x^{\tau_i(n-1)}_{\mathcal{V}\setminus i}
=[x^{\tau_i(n-1)}_{1},\ldots, x^{\tau_i(n-1)}_{i-1},x^{\tau_i(n-1)}_{i+1},\ldots, x^{\tau_i(n-1)}_{M}]^T$.
The synchronous version of (\ref{evolution}) is a special case of asynchronous updating in  (\ref{evolution}) without considering the possible information loss due to communications and different user may have different updating frequencies.
Suppose user $i$ only update and the time instance $\mathcal T_i$, then the outgoing information is
\begin{equation} \label{asyn2}
x_i^{(n)} =
\left\{ \begin{array}{l}
 \mathcal{F}_i(\bm x^{\tau_i(n-1)}_{\mathcal{V}\setminus i}), \;\;     \text{if $n\in \mathcal T_i$} \\
x_i^{(n)}, \;\;\;\;\;\;\;\;\;\;\;\;\; \text{otherwise.}
\end{array} \right.
\end{equation}
Following the same arguments in the proof of Theorem \ref{C-converge-iff}, we have the following convergence properties for the asynchronous updating.
\begin{mycorollary} \label{C-converge-iff}
With the asynchronous updating as in (\ref{asyn2}), $\bm x^{(n)}$ is convergence guaranteed  for any initial value $ \bm x^{(0)}\in \mathcal S$.
\end{mycorollary}

\section{Performance Evaluation}\label{simulation}
This section provides simulation results on the tests of the
developed  synchronous and asynchronous distributed multi-user wireless charging power allocation algorithms proposed in Section \ref{Nash}.
The related wireless charging parameters
are generated according to a  wireless charging prototype in \cite{Katabi-Multi} which charges users iPhone 4s based on MRC method.
More specifically, for different users,
the   received power $\frac{C_i}{D_i}$ varies  within
$\mathcal R = [1, 3]$W
The corresponding power transmission efficiency $h_i$,
according to the experiment results  is varies within $\mathcal H = [11\%, 19\%]$.
The input power $P$ is set to be $20$W, which is the mean of standardized wireless charging product including Duracell Powermat, Energizer Qi, LUXA2 and WiTricity WiT-500.
Note that, with charging methods based on different wireless power charging principle, such as MRC, RF energy transfer and inductive coupling \cite{inductive},  the charging efficiency $h_i$ is different.
Besides,
each user may require different
amount of energy $C_i$ to fulfill their devices and have different expected charging time $D_i$.
Thus there may be different methods to set these simulation parameters.
However, the proposed distributed multi-user charging power allocation algorithm and Nash equilibrium analysis is adapt to different charging methods and users.
For the asynchronous updating, the probability of user $i$ successfully pass $x_i^{(n)}$ to user $j$ is
$p_{i,j}$, with $p_{i, j}\leq 1$.
Therefore, $p_{i, j}$ emulate the asynchronous network due to packet loss or communicating delay.
The weighting parameter $\lambda$ is set to be $1$ without specification.

\begin{figure}[t]
\centering
\epsfig{file=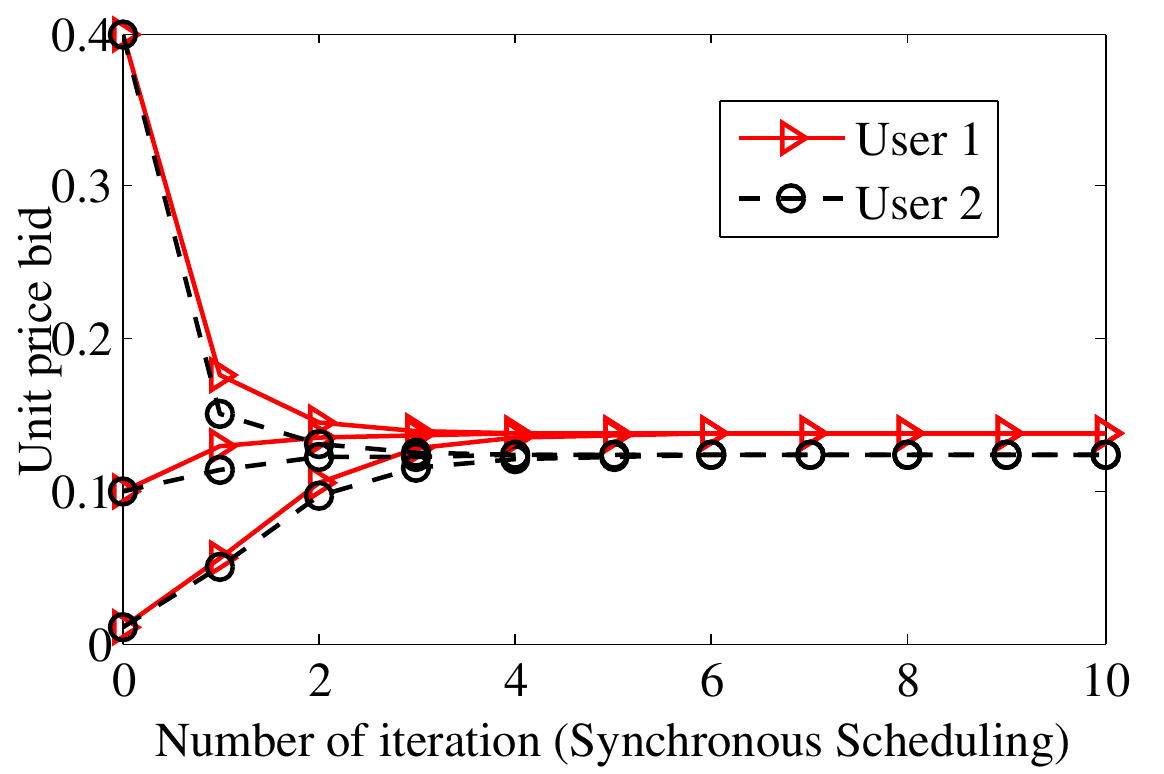, width=3in}
\caption{Convergence performance of the proposed algorithm for different initial values with synchronous scheduling.}
\label{DiffInitil}
\end{figure}

\begin{figure}
\centering
\epsfig{file=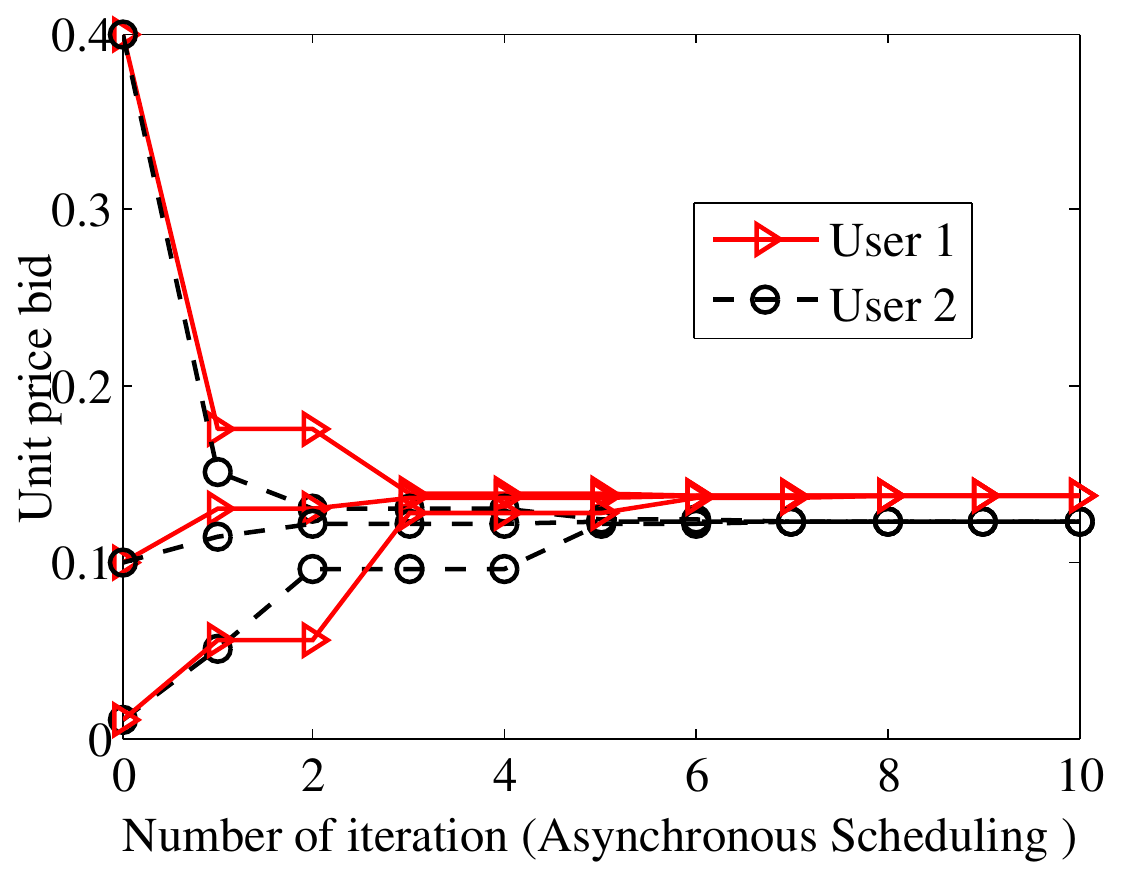, width=3in}
\caption{Convergence performance of the proposed algorithm for different initial values with asynchronous scheduling.}
\label{DiffInitil-Asyn}
\end{figure}

Fig. \ref{DiffInitil} shows the convergence behavior of the proposed
algorithms for computing the Nash equilibrium $x_i^{\ast}$.
Two users compete for the wireless power charging is considered.
$C_i/D_i$ for $i=1, 2$ are randomly picked up from  $\mathcal R $.
It is assumed there is no packet loss nor any communication delay, thus $p_{1,2}=p_{2,1}=1$.
We start the distributed synchronous updating algorithm with different values.
It shows in Fig. \ref{DiffInitil} that each user's $x_i^{(n)}$ converges to the unique Nash equilibrium within $4$ iterations.
Moreover, Fig. \ref{DiffInitil-Asyn} shows the convergence property for asynchronous scheduling with the same parameter set up but a successfully communication probability $p_{1,2}=p_{2,1}=0.8$.
It can be seen that:
The distributed
approach converges still very rapidly and after convergence
the corresponding Nash equilibrium is unique.

\begin{figure}[t]
\centering
\epsfig{file=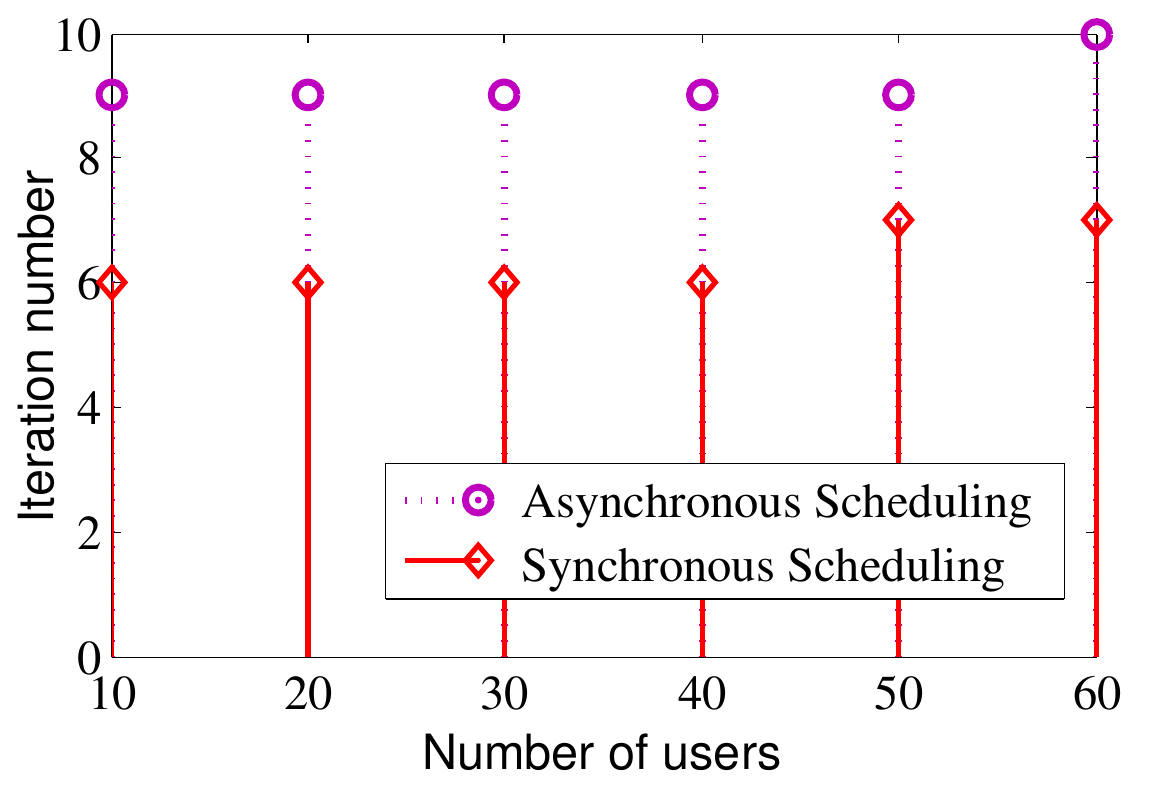, width=3in}
\caption{The iteration numbers upon convergence versus the network size  for both synchronous and asynchronous scheduling.}
\label{Iter-No}
\end{figure}

Fig. \ref{Iter-No} shows the relationship between iteration number upon convergence versus  the network size for both synchronous and asynchronous scheduling ($p_{i,j}=0.8$).
It is shown that for both cases, the
 the iteration number increases slightly when the network size increases.
The slow increase of iteration number is   due to the exponential convergence rate as shown in Theorem \ref{RateCov}.
Thus for different network sizes, the iteration number is quite small and almost the same.
Fig. \ref{Iter-No} also shows with $p_{i,j}=0.8$,  the relationship between iteration number upon convergence versus  the network size under asynchronous updating.
It can be seen that even with $20\%$ chance of packet loss between each pair of users, the asynchronous algorithm can still reach the Nash equilibrium quickly.
Moreover, the computational complexity for the distributed power allocation algorithm is quite low since
for each iteration, the computation for the unit price bid $x_i^{(n)}$
in (\ref{evolution}) only involves serval scalar multiplications.

\begin{figure}[t]
\centering
\epsfig{file=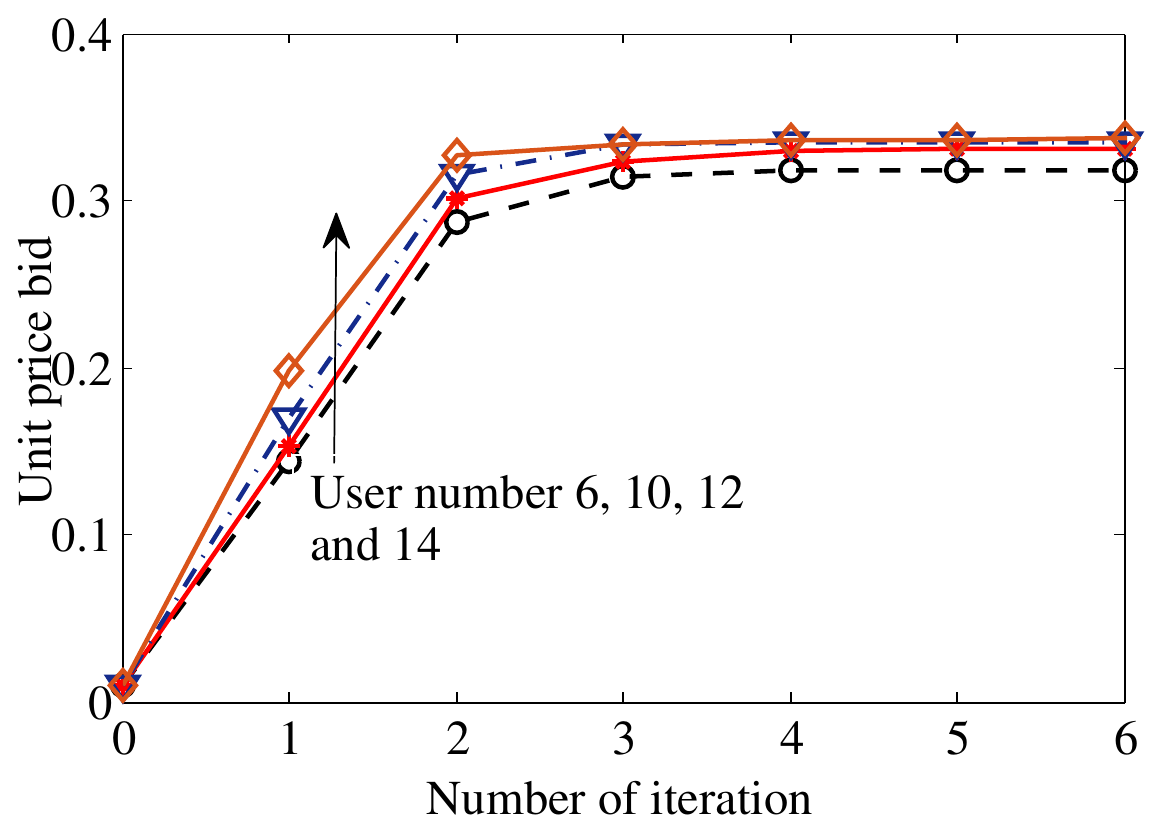, width=3in}
\caption{Convergence to the Nash equilibrium solution $x_i^{\ast}$ with respect to different number of users. (ratio = 0.7, asynchronous 0.2 success)}
\label{NashvsNo}
\end{figure}

In Figure \ref{NashvsNo}, we show the Nash equilibrium with respect to the number of users $M$.
For each case all the parameters are set to be the same for a fair comparison.
For each user, $C_i/D_i$ is set to be $1.2$.
It is demonstrated that the Nash equilibrium for each user $x_i^{\ast}$ monotonically increases as $M$ increases.
This is due to the completion among the users gradually become more severe such that the bid offered by each user at the equilibrium also increases.

\begin{figure}[t]
\centering
\epsfig{file=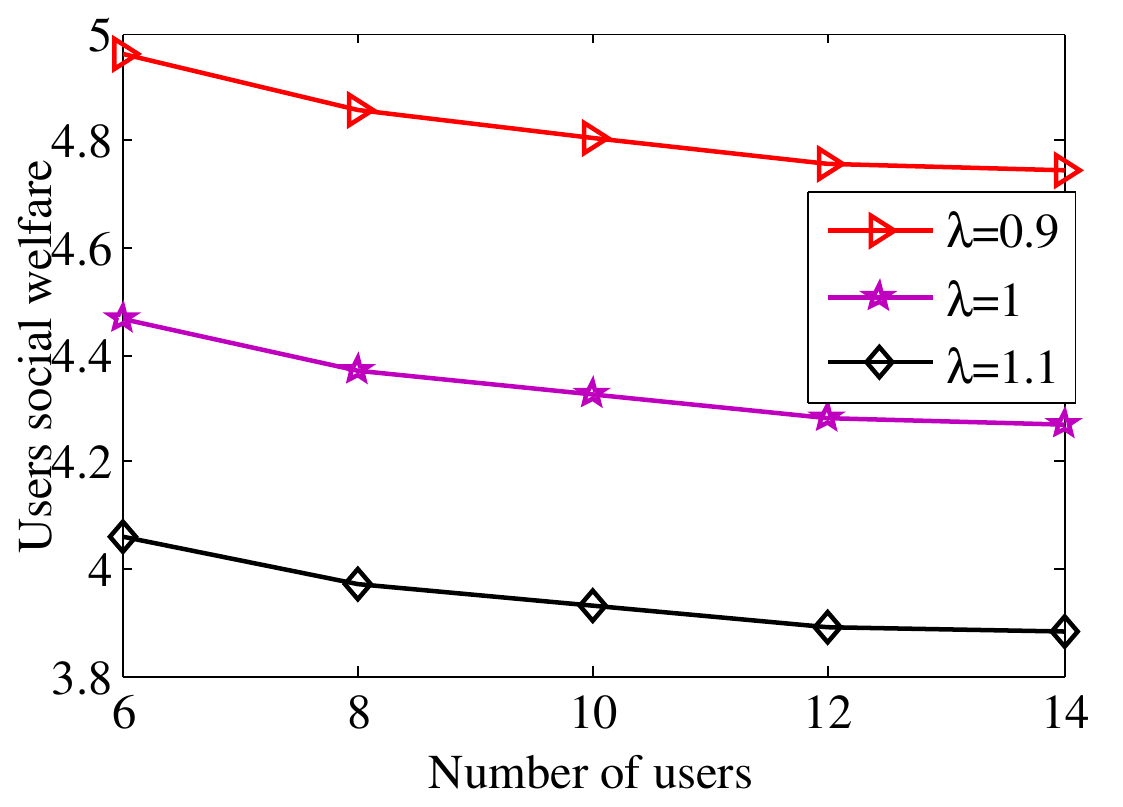, width=3in}
\caption{Users' Social welfare $\mathcal U(\bm x^{\ast})$ with respect to users number for different weighting $\lambda$.  (ratio = 0.7, synchronous 0.2 success)}
\label{SocialWelfare}
\end{figure}

\begin{figure}[]
\centering
\epsfig{file=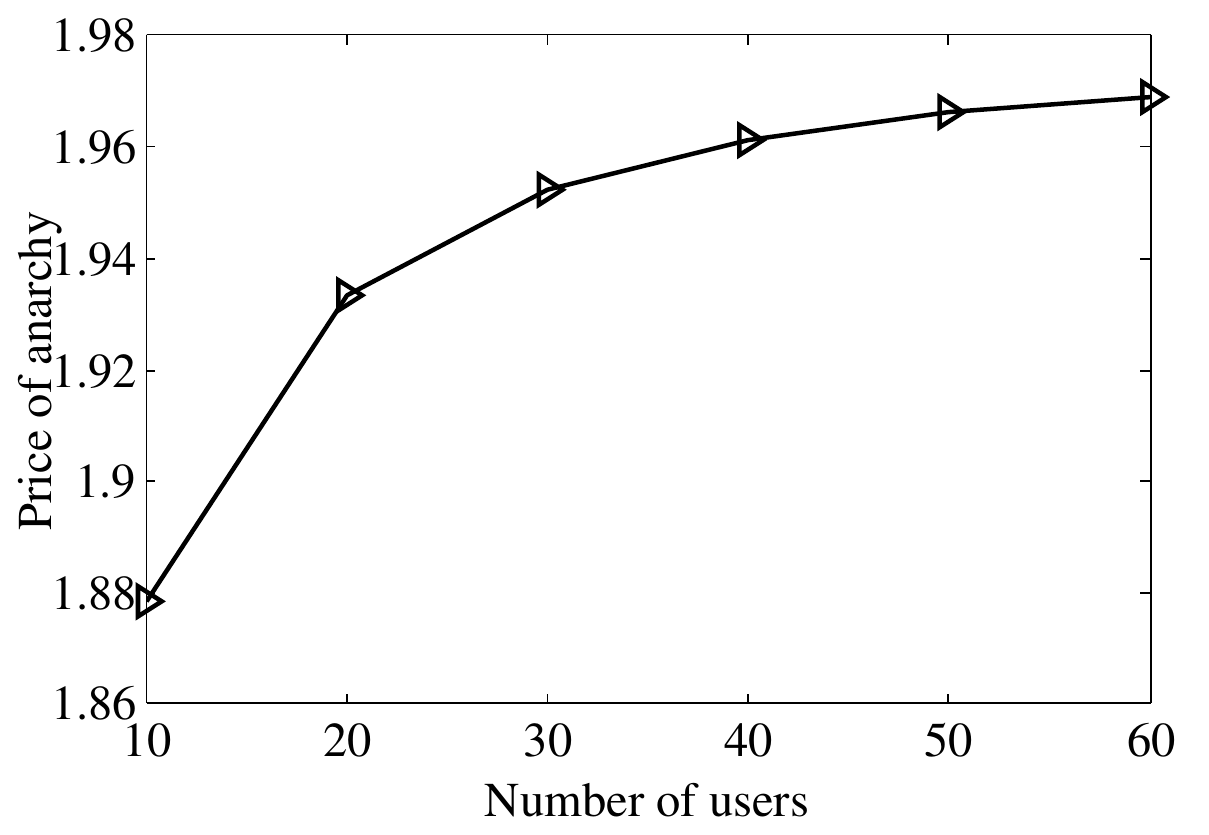, width=3in}
\caption{Price of anarchy with respect to the number of users.}
\label{PoA}
\end{figure}
Next we demonstrate the social welfare behavior of wireless charging user.
The social welfare of the users is defined as
\begin{equation}\label{utility}
\begin{split}
\mathcal{U}(\bm x^{\ast})
&= \sum_{i=1}^M\mathcal{U}_i(x^{\ast}_i, \bm x^{\ast}_{\mathcal{V}\setminus i})\nonumber\\
&= \sum_{i=1}^M
 \frac{D_i P h_ix^{\ast}_i}{C_i\sum_{j\in \mathcal{V}} x^{\ast}_j}  -
\lambda \frac{P (x_i^{\ast})^{2}}{ \sum_{j\in \mathcal{V}} x^{\ast}_j}.
\end{split}
\end{equation}
For each fixed user number,
$C_i/D_i$ are uniformly generated from $R$
and $h_i$ is uniformly generated from $D$.
Each point in the figure is an average of 1000 times simulation.
First, Figure \ref{SocialWelfare} shows that the social welfare of the users    monotonically decreases for the reason of the increasing competition among the users, which implies the increasing social welfare of the charging server.
Second, if as the portion of urgency in the utility function decreases the utility function is increases.

The efficiency of a noncooperative game is defined as the ratio of the highest value of the social welfare to the worse Nash equilibrium of the game.
By assuming users are cooperative,   the highest value of the social welfare is computed by
\begin{equation}
\bm x^{\textrm{P}} =[x_1^P,\ldots x_M^{\textrm{P}}]
= \arg\min_{\bm x}
\sum_{i=1}^M\mathcal{U}_i(x_i, \bm x_{\mathcal{V}\setminus i}).
\end{equation}
Besides, since we have shown in Theorem \ref{unique} that there is a unique Nash equilibrium in $\mathcal G$, the price of anarchy of the game $\mathcal G$ is
\begin{equation}
\textrm{PoA}
= \frac{\sum_{i=1}^M\mathcal{U}_i(x^{\textrm{P}}_i, \bm x^{\textrm{P}}_{\mathcal{V}\setminus i})}
{\sum_{i=1}^M\mathcal{U}_i(x^{\ast}_i, \bm x^{\ast}_{\mathcal{V}\setminus i})}.
\end{equation}
Figure \ref{PoA} shows that the price of anarchy increases slowly with the number of users which implies the social welfare gap between cooperative optimal and noncooperative optimal increases slowly.

\section{Conclusion}\label{conclusions}
In this paper,  a fully distributed multi-user wireless charging power allocation
scheme has been proposed.
It has taken into consideration of the impact of charging time on the power allocating problem.
The  proposed distributed algorithm
only involves limited information  exchanges among users and avoids sharing privacy information as in the centralized algorithm.
We have presented a rigorous analysis
for the existence and unique of the Nash equilibrium for both synchronous and asynchronous updating.
Moreover, we have proved that the distributed algorithm converges with a exponential rate.
Simulations have verified the theatrical analysis and shown the efficiency of the power allocation algorithm.

\appendices
\section{ }\label{A}
By taking the second derivative of $\mathcal{U}_i(x_i, \bm x_{\mathcal{V}\setminus i})$  in (\ref{utility})
with respect to $x_i$, we obtain
\begin{equation}
\begin{split}
\frac{\partial^{2}\mathcal{U}_i(x_i, \bm x_{\mathcal{V}\setminus i})}{\partial x_i^{2}}
=
&-2\frac{D_iP h_i\sum_{j\in \mathcal{V}\setminus i} x_j}{C_i(\sum_{j\in \mathcal{V}} x_j)^3}\\
&-2\lambda P
\frac{(\sum_{j\in \mathcal{V}\setminus i} x_j)^2}{C_i(\sum_{j\in \mathcal{V}} x_j)^3}.
\end{split}
\end{equation}
As $x_j>0$ for all $j\in \mathcal V$, it is clear that
${\partial^{2}\mathcal{U}_i(x_i, \bm x_{\mathcal{V}\setminus i})}/{\partial x_i^{2}}< 0$.
Therefore, the utility function $\mathcal{U}_i(x_i, \bm x_{\mathcal{V}\setminus i})$ is strictly concave,
and  the best response function, according to Definition \ref{def}, can be obtained by setting the first derivative of $\mathcal{U}_i(x_i, \bm x_{\mathcal{V}\setminus i})$ to be zero, i.e.,
\begin{equation}
\begin{split}
\frac{\partial \mathcal{U}_i(x_i, \bm x_{\mathcal{V}\setminus i})}{\partial x_i}
=
&\frac{D_iP h_i}{C_i}\frac{\sum_{j\in \mathcal{V}\setminus i}x_j}{(\sum_{j\in \mathcal{V}} x_j)^2}\\
&-\lambda P
\frac{x_i^2+ 2x_i\sum_{j\in \mathcal{V}\setminus i} x_j}{(\sum_{j\in \mathcal{V}} x_j)^2}\\
&= 0.
\end{split}
\end{equation}
After  some algebraic manipulations, the best response function for user $i$ can be readily given by
\begin{equation}
\mathcal F_i(\bm x_{\mathcal{V}\setminus i})
=
\big[(\sum_{j\in \mathcal{V}\setminus i}x_j)^2  +  K_i\sum_{j\in \mathcal{V}\setminus i}x_j \big]^{\frac{1}{2}} - \sum_{j\in \mathcal{V}\setminus i}x_j,
\end{equation}
where $K_i = \frac{D_i h_i}{\lambda C_i}$.
\section{ }\label{B}
By taking the first-order
derivative of $\mathcal{F}_i(\bm x_{\mathcal{V}\setminus i})$ in (\ref{BR0}) with respect to $x_i$, we obtain
\begin{equation}\label{1order}
\begin{split}
\frac{\partial \mathcal{F}_i(\bm x_{\mathcal{V}\setminus i})}{\partial x_i}
=
\frac{
\sum_{j\in \mathcal{V}\setminus i}x_j  +
 \frac{1}{2}K_i}
{ \big[(\sum_{j\in \mathcal{V}\setminus i}x_j)^2  +
K_i\sum_{j\in \mathcal{V}\setminus i}x_j \big]^{\frac{1}{2}} }
-
1.
\end{split}
\end{equation}
After performing some algebraic manipulations, the numerator of (\ref{1order}) can be reformulated as
\begin{eqnarray}
&&\sum_{j\in \mathcal{V}\setminus i}x_j  +
\frac{1}{2}K_i\nonumber\\
&=&
 [(\sum_{j\in \mathcal{V}\setminus i}x_j)^2
 + K_i\sum_{j\in \mathcal{V}\setminus i}x_j
 +  (\frac{1}{2}K_i)^2  ]^{1/2}.
\end{eqnarray}
Therefore,
\begin{eqnarray}
\sum_{j\in \mathcal{V}\setminus i}x_j  +
\frac{1}{2}K_i
 >
 [(\sum_{j\in \mathcal{V}\setminus i}x_j)^2
 + \frac{1}{2}K_i\sum_{j\in \mathcal{V}\setminus i}x_j
  ]^{1/2},
\end{eqnarray}
and hence, it is clear that $\frac{\partial \mathcal{F}_i(\bm x_{\mathcal{V}\setminus i})}{\partial x_i}
>0$.
Thus, $\mathcal{F}_i(\cdot)$ or equivalently, $\mathcal{F}(\cdot)$ satisfies that
$\bm x\geq \bm y$ implies
$\mathcal{F}(\bm x)\geq \mathcal{F}(\bm y)$.

Next, we will show P 1.2.
For arbitrary $\bm x \in \mathcal S$ and $\alpha<1$, it can be shown  that
\begin{equation} \label{65}
\begin{split}
&
\big[(\sum_{j\in \mathcal{V}\setminus i}\alpha x_j)^2  +
K_i\sum_{j\in \mathcal{V}\setminus i}\alpha x_j \big]^{\frac{1}{2}}
- \sum_{j\in \mathcal{V}\setminus i}\alpha x_j\\
>&
\big[(\sum_{j\in \mathcal{V}\setminus i}\alpha  x_j)^2  +
K_i\sum_{j\in \mathcal{V}\setminus i}\alpha^2 x_j \big]^{\frac{1}{2}}
- \sum_{j\in \mathcal{V}\setminus i}\alpha x_j.
\end{split}
\end{equation}
Thus, by the definition of $\mathcal{F}_i(\cdot)$, (\ref{65}) implies  $\mathcal{F}_i(\alpha\bm x_{\mathcal{V}\setminus i}) >
  \alpha \mathcal{F}_i(\bm x_{\mathcal{V}\setminus i}) = \alpha \bm x_{\mathcal{V}\setminus i}$.
Thus we have
$\mathcal F( \alpha\bm x)> \alpha\mathcal F(\bm x)$.
With similar argument, we can also show that
$\alpha^{-1}\mathcal{F}(\bm{x}) >
\mathcal{F}(\alpha^{-1} \bm{x})$.



\end{document}